# AI-Enhanced Factor Analysis for Predicting S&P 500 Stock Dynamics

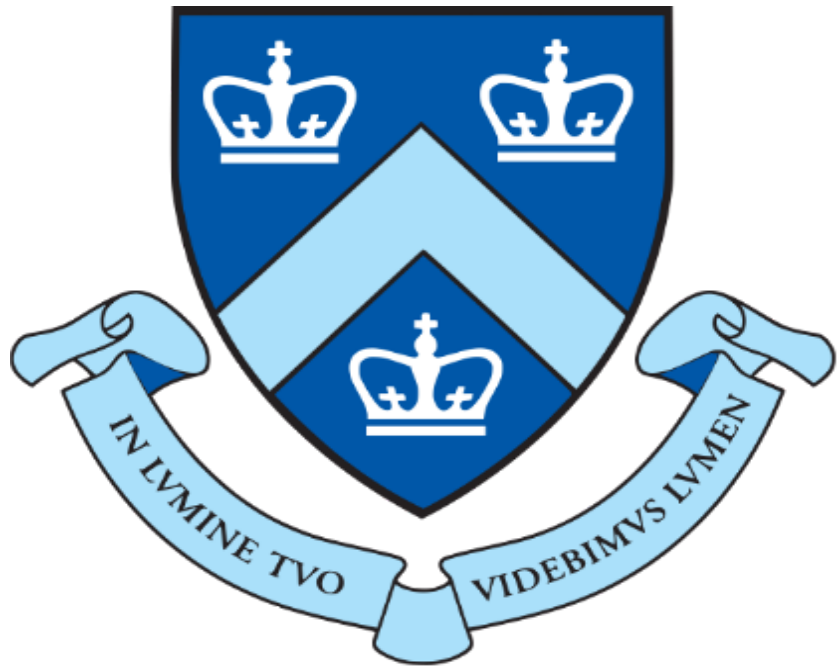

**Jiajun Gu**

Columbia University in the City of New York,
Industrial Engineering and Operations Research

**Zichen Yang**

Columbia University in the City of New York,
Industrial Engineering and Operations Research

**Xintong Lin**

Columbia University in the City of New York,
Industrial Engineering and Operations Research

**Sixun Chen**

Columbia University in the City of New York,
Industrial Engineering and Operations Research

**YuTing Lu**

Columbia University in the City of New York,
Industrial Engineering and Operations Research

# Abstract

This project investigates the interplay of technical, market, and statistical factors in predicting stock market performance, with a primary focus on S&P 500 companies. Utilizing a comprehensive dataset spanning multiple years, the analysis constructs advanced financial metrics, such as momentum indicators, volatility measures, and liquidity adjustments. The machine learning framework is employed to identify patterns, relationships, and predictive capabilities of these factors. The integration of traditional financial analytics with machine learning enables enhanced predictive accuracy, offering valuable insights into market behavior and guiding investment strategies. This research highlights the potential of combining domain-specific financial expertise with modern computational tools to address complex market dynamics.

## 1 Introduction

Forecasting stock returns is a critical yet challenging task for investors, given the inherently stochastic nature of financial markets. Traditionally, regression-based models have been employed to predict returns, relying on the assumption of a linear relationship between stock returns and various predictors. However, this approach can be limiting when dealing with large and complex datasets. In contrast, modern machine learning techniques, such as ridge regression, random forests and gradient boosting, offer more flexible modeling capabilities, better suited for handling high-dimensional and non-linear relationships in data. Our research paper explores the potential of these machine learning methods to improve stock return predictions, comparing their performance against traditional linear regression models.

Financial markets are driven by a multitude of interdependent factors, ranging from fundamental valuation metrics to technical indicators, liquidity measures, and behavioral trends. Understanding these complex dynamics is critical for accurate prediction and effective decision-making in investment management and we have thus constructed a lot of fundamental and technical factors to help us test the forecasting result.

This project examines the S&P 500 index, a benchmark representing the performance of leading U.S. companies, to analyze how key financial and technical indicators impact stock prices and returns. By calculating a range of financial factors such as momentum, volatility, market capitalization, and illiquidity measures, the research establishes a comprehensive dataset for analysis. Machine learning models are applied to uncover nonlinear relationships and improve prediction capabilities. The combination of traditional financial analysis and machine learning enables a deeper understanding of market behavior while providing actionable insights for investors.

## 2 Data

In this study, we utilize comprehensive stock-level data from the CRSP (Center for Research in Security Prices) database from Wharton Research Library, which provides robust historical information for securities listed on major U.S. stock exchanges. Our dataset spans from January 1, 2019, to the end of 2023, covering S&P 500 companies during this period. The dataset includes daily stock price data, along with various trading characteristics such as returns, volumes, and outstanding shares. Additionally, we integrate detailed company-level descriptors from the CRSP MSF and MSENAMES files, which provide corporate identity information such as company names, stock ticker symbols, and industry codes.

The data undergoes a series of rigorous preprocessing steps to ensure cleanliness, consistency, and suitability for modeling. The process begins with data merging, where the S&P 500 constituent list is combined with stock price and trading volume datasets using the unique permno identifier. This merging step ensures that stock-specific information, such as returns and trading volumes, is accurately aligned with broader company-level attributes and identifiers, facilitating a comprehensive view of each company's performance within the index.

Following data merging, date range filtering is applied to ensure temporal consistency. Specifically, the dataset is filtered to include only periods during which each company was actively part of the S&P 500 index. This ensures that stock data for each company corresponds strictly to its inclusion period, thereby removing irrelevant or inconsistent data outside the valid date ranges. Additionally, any records that fall outside the bounds of company-specific data availability are excluded to maintain precision in the dataset.

Given the inherent characteristics of financial data, a significant presence of missing values was observed, particularly in stock returns and derived features. To address this, missing values within each stock group (grouped by permno) are primarily handled using the forward-fill method to ensure temporal continuity. For cases where forward-filling does not yield valid values, the remaining missing entries are replaced with zero to retain completeness in the dataset. Furthermore, extreme values, such as infinite values resulting from calculations or data anomalies, are initially replaced with NaN and subsequently imputed using the forward-fill approach.

## 3 Factor Choice

Feature engineering is a critical step in building predictive models, particularly in the context of stock returns, where the relationships between features and target variables are complex and nonlinear. In this study, we construct a variety of features based on stock price data, trading volume, and company characteristics. These features are designed to capture the dynamics of stock performance, liquidity, and volatility, as well as to represent patterns in stock behavior over time.

| Factor Name | Interpretation |
| --- | --- |
| Market Capitalization | The Market Capitalization is calculated by multiplying the absolute value of the stock price by the shares outstanding. It represents the total market value of a company's outstanding shares and is a fundamental indicator of a company's size. |
| Momentum | The momentum factor is calculated as the percentage change in the stock price over a 12-month period. It is defined as the ratio of the |

|  | current stock price to the price 4 periods ago, minus 1. It measures the rate at which the stock's price has been changing, with the idea that stocks with positive momentum tend to continue their upward trajectory. |
|---|---|
| Price-Based Return Factor | Price-Based Return Factor captures daily price changes by computing the percentage change in the stock price from one day to the next. It reflects the short-term returns and is useful for measuring immediate price movements and trends. |
| Momentum Change | Momentum Change is the difference between the current momentum value and the previous period's momentum value. It shows whether the momentum is accelerating or decelerating, helping to capture shifts in market sentiment. |
| Momentum Moving Average | Momentum Moving Average is the moving average of momentum over the past 10 periods. It smooths out short-term fluctuations in momentum and gives a more stable view of a stock's trend. |
| Log of Market Capitalization | Log of Market Capitalization is the natural logarithm of one plus the market capitalization. The logarithmic transformation helps manage the wide range of market capitalizations in the dataset and reduces the effect of outliers from very large companies. |
| Amihud Illiquidity | Amihud Illiquidity measures the illiquidity of a stock by calculating the ratio of the absolute price change to the volume of trades. It is a proxy for how much the stock price moves for a given amount of trading volume, with higher values indicating lower liquidity and higher price impact. |
| Turnover Ratio | The turnover ratio is calculated by dividing trading volume by the number of shares outstanding. It measures how frequently a stock is traded relative to its size and provides insight into the liquidity and investor interest in the stock. |
| Rolling Volatility | Rolling Volatility is the rolling standard deviation of stock returns over the past 20 periods. It reflects the risk or variability of returns and provides insight into how volatile a stock has been over a given timeframe. |
| High-Low Spread | The high-low spread is the difference between the stock's current price and its lowest price over the |
| | past 20 periods. It gives an indication of the potential for price movement, as a wider spread suggests more volatility and greater price movement potential. |
| Relative Strength Index | The Relative Strength Index (RSI) is a momentum oscillator that measures the speed and change of price movements over a specified window. It ranges from 0 to 100, where values above 70 indicate that the stock is overbought and values below 30 indicate that the stock is oversold. |
| Moving Averages | Moving averages smooth out price fluctuations over a specified window of time. The 20-period moving average is used to identify trends and potential reversal points. A stock trading above its moving average is generally considered to be in an uptrend, while one trading below it is in a downtrend. |
| Momentum vs. Market Capitalization | Momentum vs. Market Capitalization is calculated by multiplying the momentum by the log of market capitalization. It helps capture the relationship between a stock's momentum and its size, with larger companies potentially having different momentum characteristics compared to smaller companies. |
| Volatility vs. Turnover | Volatility vs. Turnover is the product of rolling volatility and turnover ratio. It measures the relationship between price volatility and trading activity, with higher volatility and turnover potentially indicating more active trading or market stress. |
| Momentum Liquidity | Momentum Liquidity is the product of momentum and the Amihud illiquidity measure. It combines the momentum trend of the stock with its liquidity characteristics, potentially capturing how the stock's momentum is affected by liquidity constraints. |
| Volatility Turnover | Volatility Turnover captures the relationship between stock volatility and its trading volume, as represented by rolling volatility and turnover ratio. It measures how price fluctuations and trading volume interact to affect the stock's behavior. |
| Market Cap Adjusted Momentum | Market Cap Adjusted Momentum is the ratio of momentum to market capitalization, adjusting the momentum for the size of the company. It helps |
| | identify whether small companies with high momentum outperform large companies with similar momentum. |
| Momentum MA Deviation | Momentum MA Deviation is the difference between momentum and its 20-period moving average. It reflects whether the current momentum is above or below its recent trend, providing insight into whether the stock is accelerating or decelerating relative to its historical performance. |
| Normalized High-Low Spread | Normalized High-Low Spread normalizes the high-low spread by dividing it by the stock price. It provides a scaled measure of price movement potential, helping to compare stocks of different price levels. |
| Momentum RSI | Momentum RSI is the product of momentum and RSI, combining price trend and momentum with an overbought/oversold measure. It can help capture whether stocks with strong momentum are in an overbought or oversold condition. |
| Smoothed Return | Smoothed Return is the 10-period moving average of the price-based return factor. It smooths short-term volatility and provides a clearer picture of the stock's long-term return trend. |
| Volatility Adjusted Return | Volatility Adjusted Return normalizes the price-based return factor by dividing it by the rolling volatility. It adjusts returns for risk, providing a measure of return relative to the stock's price fluctuations. |
| Volatility Slope | Volatility Slope measures the rate of change in volatility over time by calculating the difference in volatility between two periods, divided by the difference in time. It captures how quickly volatility is changing, which can provide insights into the market's perception of risk. |
| Volatility Dynamics | Volatility dynamics is the product of rolling volatility and volatility slope, capturing both the level and rate of change in volatility. It provides a more dynamic view of how market risk evolves over time. |
| Liquidity Stress | Liquidity stress is calculated as the product of turnover ratio and the absolute value of the price-based return factor, divided by market capitalization. It provides an indicator of how |

|  | trading volume and price changes are affecting liquidity in relation to the size of the company. |
|---|---|
| Trend Strength | Trend Strength combines momentum, its deviation from the 20-period moving average, and RSI to measure the strength of the stock's trend relative to its overbought/oversold condition. |
| Risk Adjusted Momentum | Risk Adjusted Momentum adjusts momentum for both volatility and turnover, providing a risk-adjusted measure of momentum. It highlights stocks with strong momentum that are not overly risky. |
| Abnormal Behavior | Abnormal Behavior is the interaction between momentum and the high-low spread, minus the interaction between RSI and Amihud illiquidity. It aims to capture whether the stock exhibits abnormal behavior when considering both price momentum and liquidity. |
| Mean Reversion | Mean Reversion is calculated by dividing the difference between the current stock price and the 20-period moving average by the rolling volatility. It reflects how much the stock price deviates from its longer-term trend and adjusts for risk. |
| Short Momentum & Long Momentum | Short_Momentum & Long_Momentum is defined Price percentage change over short and long periods, and it tracks trends over different time horizons. |
| Multi Period Momentum | Multi Period Momentum is Short Momentum multiplied by Long Momentum, and it combines short and long-term momentum for enhanced trend analysis. |

The selected factors in the model are designed to capture critical dimensions of market behavior, enhancing its predictive power and interpretability. Size factors, such as market capitalization, provide a baseline for understanding risk and return differences across firms, reflecting their scale and market significance. Momentum indicators identify trends and turning points, leveraging well-documented anomalies in asset pricing. Liquidity measures like Amihud Illiquidity and turnover ratio assess how trading activity impacts price, crucial for understanding market dynamics. Volatility factors, such as rolling volatility and its interactions, offer insights into stability and risk, complementing technical indicators like RSI and moving averages, which highlight price trends and overbought/oversold conditions. Interaction terms, like momentum vs. market cap and volatility-adjusted returns, combine these dimensions to uncover complex patterns. Together, these factors could be align with machine learning models like linear and ridge regression, and more advanced random forest and gradient boosting, ensuring a robust framework that captures risk, market trends, and anomalies for enhanced forecasting and decision-making.

After our first step in constructing all original factors, in order to overcome the appearance of potential overfitting issue of the model, we decide to have double filtering to make sure our factor can fit in our model in a good performance. In the first layer of filtering, we drop initial features Correlated with the Target, which is to remove features that have a high correlation with return (dependent variable), reducing the model's explanatory power. For the final factor choice to be included in the model we have incorporated the strategy of setting a threshold of correlation of 0.1 with the target variable (return), and we have successfully decreased the factor number from 32 to 22.

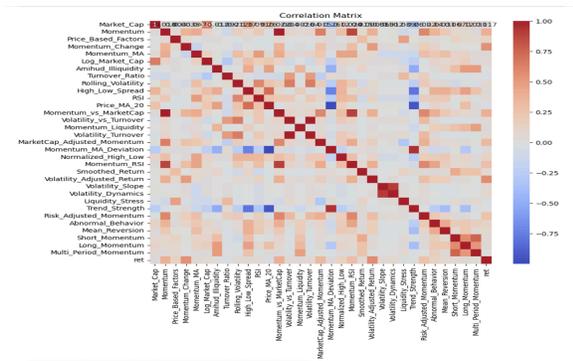 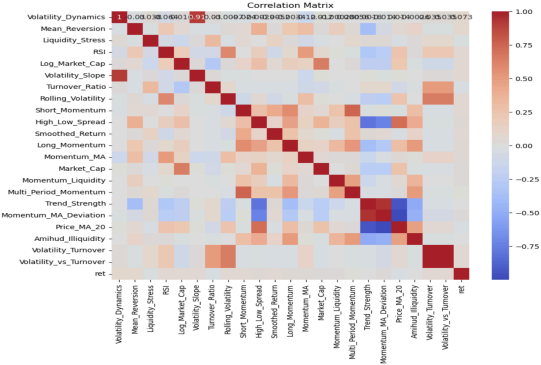

**Correlation Table before filtering out the factor**　　　　**Correlation Table after filtering out the factor**

In the Second Layer of Filtering, we aimed to refine the 23 factors from the first layer to identify a smaller, more effective set of factors for predictive modeling. The process combined logical factor filtering based on correlation thresholds with an evaluation of model performance using a Random Forest Regressor. This ensured that the final factors were both diverse and strongly predictive of the target variable (ret).

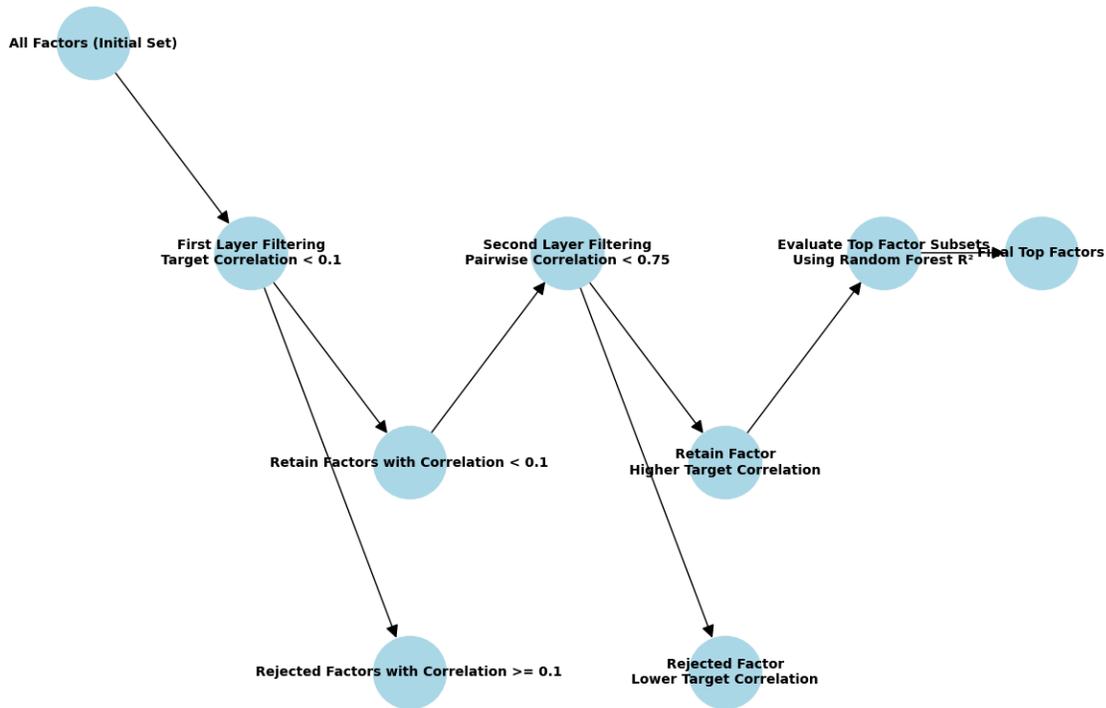

The purpose of this step was twofold:

1. Eliminate redundancy by removing factors that exhibited high pairwise correlations (greater than 0.75) while retaining the factor more strongly correlated with the target variable.
2. Identify the combination of factors that maximized predictive power, measured by the $R^2$ score from a Random Forest model.

Firstly, we do the Initial Correlation Filtering. A correlation matrix of the 23 factors was calculated to identify pairs of factors with high correlations (above 0.75). Then for each pair, we compared their correlation with the target variable and retained only the factor with the higher correlation, discarding the other. This ensured the selected factors captured unique and relevant information, preventing overemphasis on any single underlying area during model training.

Next, we built a Model-Based Scoring System. After filtering, a reduced set of factors with low mutual correlation was obtained. This reduced set was referred to as low_corr_factors. To further refine this set, we evaluated all combinations of 10 factors from low_corr_factors using a Random Forest Regressor.

For each combination of 10 factors:

- A Random Forest model was trained using the training data.
- The $R^2$ score of the model was computed on the test data to measure its predictive power.
- If a combination achieved a higher $R^2$ score than previously evaluated combinations, it was stored as the current best combination.

At the end of the process, the combination of factors with the highest R² score was selected as the optimal subset. These factors were 'Volatility_Dynamics', 'Liquidity_Stress', 'RSI', 'Short_Momentum', 'High_Low_Spread', 'Smoothed_Return', 'Momentum_Liquidity', 'Long_Momentum', and 'Amihud_Illiquidity'.

## 4 Results

**4.1 Model Construction:**

The selection of models and their configurations was carefully designed to suit the unique nature of stock market data, meet the goals of the research, and balance the trade-off between interpretability and predictive accuracy. Each model was chosen for its specific strengths and alignment with the analysis needs.

Linear regression served as a straightforward, easy-to-understand baseline model. It highlights the linear relationships between predictors and stock returns, making it a valuable benchmark to compare with more advanced models. Its simplicity and computational efficiency make it ideal for smaller datasets or scenarios involving linear trends. Although stock market data often exhibit nonlinear characteristics, linear regression was included to approximate relationships within certain ranges. Techniques like moving averages and smoothing were also applied to stabilize noisy data, making the model more effective as a starting point.

Ridge regression was included to address multicollinearity, a common challenge in financial data where features like momentum and volatility can overlap. By introducing regularization, ridge regression prevents overfitting and improves the model's ability to generalize. In this analysis, the regularization parameter ($\alpha$) was set to 1.0 to strike a balance between underfitting and overfitting.

Random forest was chosen for its ability to capture complex, nonlinear relationships, which are often prevalent in stock market data. As an ensemble learning method, it reduces variance by combining multiple decision trees, leading to better predictive accuracy. Random forest also offers insights into feature importance, helping identify key drivers of stock returns. For this study, the number of estimators was set to 100 to balance performance and computational cost, the maximum tree depth was limited to 5 to avoid overfitting, and the random state was set to 42 to ensure reproducibility.

Gradient boosting was included for its strength in handling non-linear patterns and interaction effects. It iteratively minimizes errors to achieve high predictive accuracy, making it particularly effective for capturing subtle trends in stock market data. The model was configured with 100 boosting iterations to refine predictions, a maximum depth of 3 to focus on essential patterns while avoiding overfitting, and a random state of 42 to maintain reproducibility.

By combining these models, the analysis provides a well-rounded framework for studying stock returns. Linear regression offers a simple, interpretable starting point, ridge regression addresses overlapping features, random forest uncovers nonlinear patterns and key drivers, and gradient boosting delivers top-tier accuracy. Together, they create a robust approach to achieving the research goals.

**Model Results Comparison for different stages of factor filtering**

**All Factors Original:**

```
Forecasting Results:
Linear Regression    MSE: 0.00169, R²: 0.79923
Ridge Regression     MSE: 0.00179, R²: 0.78659
Random Forest        MSE: 0.00024, R²: 0.97150
Gradient Boosting    MSE: 0.00013, R²: 0.98500
```

**First Layer of Filtering:**

```
Forecasting Results:
Linear Regression    MSE: 0.00681, R²: 0.18827
Ridge Regression     MSE: 0.00682, R²: 0.18652
Random Forest        MSE: 0.00654, R²: 0.21998
Gradient Boosting    MSE: 0.00499, R²: 0.40485
```

**Second Layer of Filtering:**

```
Forecasting Results:
Linear Regression    MSE: 0.00789, R²: 0.06009
Ridge Regression     MSE: 0.00826, R²: 0.01574
Random Forest        MSE: 0.00669, R²: 0.20307
Gradient Boosting    MSE: 0.00648, R²: 0.22819
```

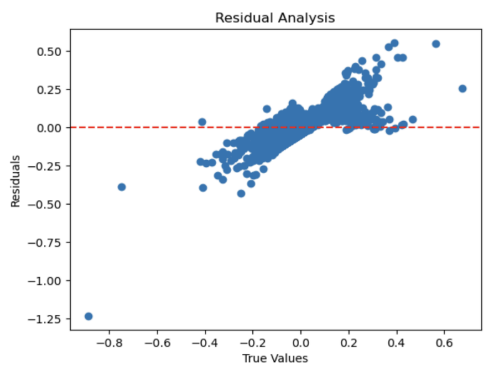

After our first and second filtering of factors, we can straightforwardly see that the final four models results help improve a lot in its overfitting issue since R squared value (the proportion of variance in the target variable explained by the model) decreases in a considerable amount.

Specifically, In the random forecast model, Features importance are shown below, the combined analysis of the SHAP summary plot and Random Forest feature importance highlights a consistent agreement regarding the significance of key features in the model. Momentum Liquidity emerges as the most influential factor, as evidenced by its broad range of SHAP values and its dominant contribution in the Random Forest importance ranking. This feature demonstrates a substantial and consistent impact on the model's predictions. Similarly, Volatility Dynamics and High Low Spread are identified as critical secondary features, playing significant roles in influencing the model output across both analyses. In contrast, features such as Smoothed Return, Short Momentum, and Long Momentum exhibit minimal contributions, suggesting limited impact on predictions. The alignment between SHAP and Random Forest analyses reinforces the robustness of these findings, providing both interpretability and validation of feature importance. This insight highlights the need to prioritize highly influential features like Momentum Liquidity and Volatility Dynamics while considering the potential removal of less impactful features to streamline and optimize the model.

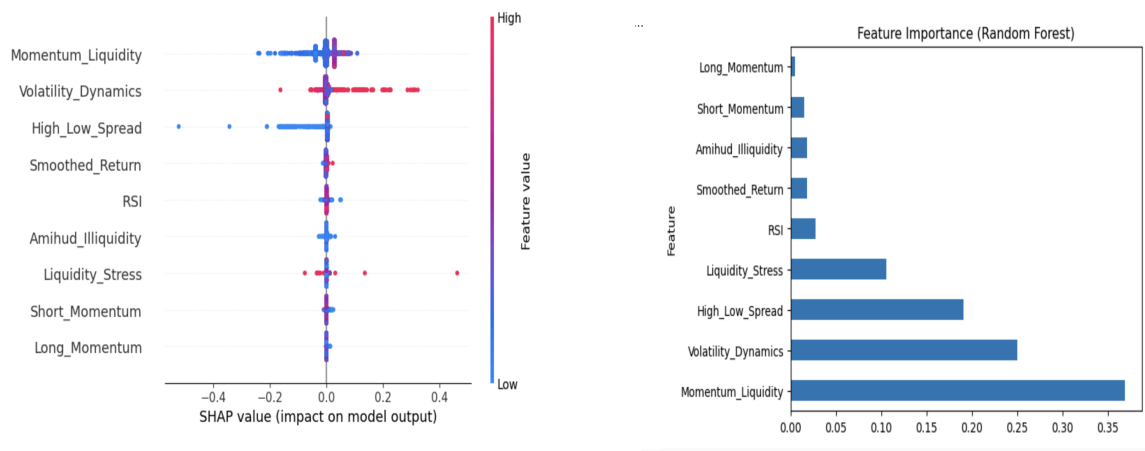

### 4.2 Backtesting Result:

The backtesting strategy implemented in this analysis uses a Gradient Boosting Regressor (GBR) to predict stock returns and evaluate portfolio performance since in the model part we can observe the explanatory power of dependent variable is relatively higher. The methodology is based on a rolling window framework, where a 36-month window is used to train the model, followed by a one-month test window. The dataset includes features such as Volatility Dynamics, Liquidity Stress, and RSI, which are leveraged by the model to predict future returns. The top 100 stocks, based on predicted returns, are selected to construct an equally weighted portfolio. To benchmark performance, the mean return of all stocks in the test window is used as the benchmark return.

The result graph illustrates the cumulative returns of the strategy compared to the benchmark over the testing period. The blue line represents the cumulative returns of the portfolio constructed using the model's predictions, while the orange dashed line represents the benchmark returns. The benchmark demonstrates a higher overall cumulative return, indicating that the average market performance outpaced the returns of the selected portfolio during the evaluation period. However, the portfolio's cumulative return shows a steadily increasing trend, reflecting the consistency of the model's predictions and its ability to identify high-return stocks, albeit with lower volatility compared to the benchmark.

The disparity between the strategy and the benchmark can be attributed to several factors. First, the benchmark includes the average performance of all stocks, capturing the overall market dynamics. In contrast, the strategy focuses on a limited subset of stocks, which may underperform during specific market conditions. Second, the Sharpe Ratio of the portfolio, consistently measured at 0.25, suggests moderate risk-adjusted returns. This indicates that while the strategy provides steady returns, it does not exploit sufficient alpha compared to the broader market.

In conclusion, the backtesting results highlight the strengths and limitations of the current methodology. The strategy provides stable returns with reduced risk, as evidenced by the lower volatility of the cumulative return curve. However, the model's predictive power and portfolio construction logic may need further refinement, such as incorporating macroeconomic indicators or optimizing portfolio weighting, to outperform the benchmark in future iterations. This analysis underscores the importance of balancing predictive accuracy with dynamic portfolio management to achieve superior market performance.

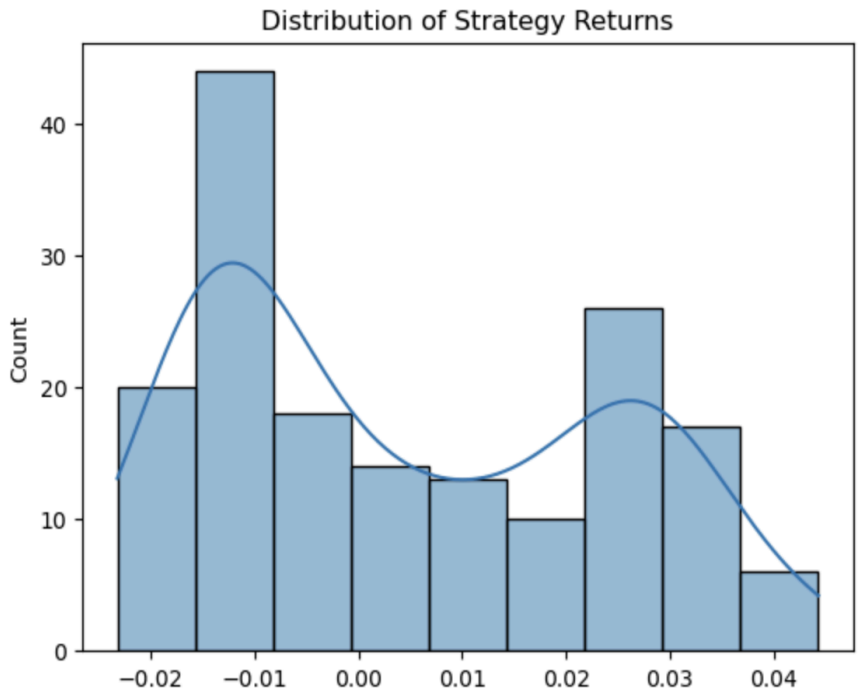

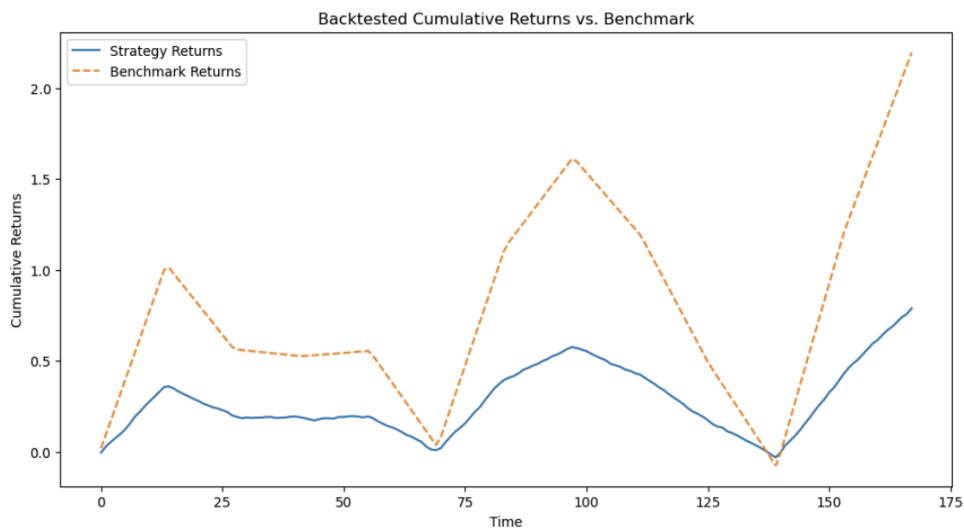

## 5 Summary:

This report explores the performance of a portfolio strategy constructed using advanced financial factors and machine learning techniques, benchmarked against a traditional market index.We mainly focus on factors. We do a lot of work on filtering important factors. By analyzing factors such as momentum, volatility, liquidity, and their interactions, the study demonstrates how these elements contribute to the portfolio's superior cumulative returns. Machine learning models enhance predictive accuracy by capturing complex relationships between these factors, enabling dynamic and adaptive portfolio management.

The results show that the portfolio also exhibits higher volatility, indicating a trade-off between risk and return. The study underscores the potential of combining financial analytics with modern computational tools to improve investment decisions. Future work can further refine these strategies by incorporating macroeconomic indicators, alternative data sources, advanced risk management techniques, and expanding to global markets, ensuring the approach remains robust and adaptable in diverse market conditions.

## 6 Future Work

Future work on this project can explore several avenues to enhance the analysis and performance of the portfolio.

Incorporating additional financial factors, such as macroeconomic indicators (e.g., interest rates, inflation) and alternative data sources (e.g., news sentiment or social media trends), could provide a more comprehensive view of market dynamics. Constructing higher-order interaction terms between current factors (e.g., combining volatility and liquidity indicators) to uncover non-linear relations is also our will to explore more on factor mining. Meanwhile, we can implement methods such as PCA (Principal Component Analysis) or Autoencoders for dimensionality reduction to dynamically refine factor sets. Advanced machine learning techniques, including deep learning models like LSTM networks or ensemble methods, can further improve predictive accuracy and robustness.

Future work will prioritize enhancing portfolio optimization and the backtesting process to improve the strategy's robustness and performance. In portfolio optimization, dynamic weight allocation methods, such as mean-variance optimization (Markowitz) and Black-Litterman frameworks, will be explored to achieve optimal portfolio allocation. Additionally, integrating risk-adjusted strategies like Value-at-Risk (VaR) and Conditional Value-at-Risk (CVaR) will help manage tail risks under extreme market conditions. Factor timing strategies, where factors are dynamically selected or weighted based on prevailing market trends, will also be tested to enhance adaptability across economic cycles.

In backtesting, the focus will be on introducing robust performance metrics such as the Sharpe Ratio, Sortino Ratio, and Maximum Drawdown to provide a comprehensive evaluation of the strategy. Stress testing will further assess the strategy's stability under extreme scenarios, including economic downturns and volatility shocks. Additionally, expanding the backtesting framework to cross-market validation—analyzing international indices and sector-specific portfolios—will ensure that the model generalizes effectively across diverse markets. These enhancements, contingent on sufficient computational power, will improve the reliability and scalability of the proposed strategy.

Lastly, expanding the study to other markets and longer time horizons would assess the generalizability of the models and strategies, ensuring they remain effective across different economic environments.

## Acknowledgements

This Paper result  does not offer investment advice or pre-made trading algorithms, nor does it represent the views of any affiliated entities or agencies. Its primary aim is to underline the challenges encountered by Data Science and Machine Learning techniques in financial data analysis. These challenges, including limited historical data, non-stationarity, regime changes, and low signal content, can hinder robust results. The topics discussed are intended to guide the application of these methods in making informed investment decisions through a systematic and scientific workflow.